\documentstyle[12pt,epsf,aasms4]{article}

\received{August 13, 1999}
\accepted{October 7, 1999}

\begin{document}

\font\fortssbx=cmssbx10 scaled \magstep2
\hbox to \hsize{
\hbox{\fortssbx University of Wisconsin - Madison}
\hfill$\vtop{\hbox{\bf MADPH-99-1134} 
                \hbox{\bf astro-ph/9908138}
                \hbox{August 1999}}$}

\vspace{.5in}

\title{Neutrino Event Rates from Gamma Ray Bursts}
\author{F. Halzen and D.W. Hooper}
\affil{Department of Physics, University of Wisconsin, Madison, WI 53706}

\begin{abstract}
We recalculate the diffuse flux of high energy neutrinos produced by Gamma Ray Bursts (GRB) in the relativistic fireball model. Although we confirm that the average single burst produces only $\sim 10^{-2}$ high energy neutrino events in a detector with $1\rm~km^2$ effective area, i.e.\ about 10 events per year, we show that the observed rate is dominated by burst-to-burst fluctuations which are very large. We find event rates that are expected to be larger by one order of magnitude, likely more, which are dominated by a few very bright bursts. This greatly simplifies their detection.
\end{abstract}


\newpage

\section{High Energy Neutrinos from Relativistic Fireballs}

The evidence has been steadily accumulating that GRB emission is the result of a relativistically expanding fireball energized by a process involving neutron stars or black holes (\cite{piran}). In the early stage, the fireball cannot emit efficiently because the radiation is trapped due to the very large optical depth. The fireball's energy is dissipated in kinetic energy until it becomes optically thin and produces the observed display. This scenario can accommodate the observed energy- and time scales provided the bulk Lorentz factor $\gamma$ of the expanding shock is $\sim300$.

The production of high energy neutrinos is anticipated: protons, accelerated in the kinetic phase of the shock, interact with photons producing charged pions which are the parents of neutrinos (\cite{waxman}). Standard particle physics and fireball phenomenology are sufficient to compute the neutrino 
flux (\cite{waxman}; \cite{halzen}) as well as the observed rates in high energy neutrino telescopes.

The observation of GRB neutrinos over a cosmological baseline has scientific potential beyond testing the ``best-buy" fireball model: the observations can test with unmatched precision special relativity and the equivalence principle, and study oscillating neutrino flavors over the ultimate baseline of $z \simeq 1$.

The anticipated neutrino flux traces the broken power-law spectrum observed for the photons, which provide the target material for neutrino production:
\begin{eqnarray}
\phi &=& \frac{A}{E_{B}E}  \mbox{\quad for }  E<E_{B} \\[3mm]
\phi &=& \frac{A}{E^{2}}  \mbox{\qquad for }  E>E_{B}\,,
\end{eqnarray}
where $A$ is a normalization constant which is determined from energy considerations and $E_{B}\approx 700$~TeV (\cite{halzen}). The total energy in GRB neutrinos is given by
\begin{equation}
 F_{\rm tot} = \frac{c}{4\pi}\frac{1}{2}f_{\pi} \, t_{\rm hubble} \, \dot\rho_E\,, 
\end{equation}
where $f_{\pi}$ is the fraction of proton energy that goes into pion production, $t_{\rm hubble}$ is 10 Byrs and $\dot\rho_E$ is the injection rate into protons accelerated to high energies in the kinetic fireball. This is a critical parameter in the calculations; it can be fixed, for instance, by assuming that GRB are the source of the highest energy cosmic rays beyond the ankle (\cite{waxmanprime}) in the spectrum, or $\dot\rho_E \simeq 4 \times 10^{44} \rm \, erg~Mpc^{-3}\,yr^{-1}$ (\cite{halzen}).  The factor $f_{\pi}$ represent the fraction of total energy going into pion production. It is calculated by known particle physics and is of order 15\% (\cite{waxman}; \cite{halzen}). We remind the reader that these assumptions reproduce the observed average photon energy per burst if equal energy goes into the hadronic and electromagnetic component of the fireball (\cite{waxman}).  

Now, normalizing the flux for this total energy:
\begin{equation}
 F_{\rm tot} = \frac{A}{E_{B}} \int_{E_{\rm min}}^{E_{B}} dE + A\int_{E_{B}}^{E_{\rm max}}\frac{dE}{E}\,. 
\end{equation} 
Approximating $E_{B}-E_{\rm min} \simeq E_{B}$, the integration constant is found to be $1.20 \times 10^{-12}\rm\,TeV\,cm^{-2}\,s^{-1}\,sr^{-1}$.
The quantities $f_\pi$ and $E_B$ are calculated using Eqs.~(4) and (5) in Waxman and Bahcall (1997). The GRB parameters entering these equations are chosen following Halzen (1998) and assumed to be independent of $\gamma$. We assumed $E_{\rm max} = 10^7$~TeV.

Given the neutrino flux $\phi$, the event rates in a neutrino detector are obtained by a straightforward method (\cite{halzen}):
\begin{equation}
 N = \int \phi(E) P_{\nu\rightarrow\mu}dE \,,
\end{equation} 
where $P_{\nu\rightarrow\mu}$ is the detection probability for neutrinos of muon flavor. It is determined, as a  function of energy, by the neutrino cross sections and the range of the secondary muon (\cite{pr}). We obtain an observed neutrino flux of order 10~yr$^{-1}\rm\, km^{-2}$ assuming $10^3$ bursts per year.  This result is somewhat lower but not inconsistent with those obtained previously.

One should keep in mind that neutrinos can also be produced, possibly with higher fluxes, in other stages of the fireball, e.g. when it expands through the interstellar medium (\cite{katz}).

\section{Burst-to-Burst Fluctuations}

We here want to draw attention to the fact that this result should not be used without consideration of the burst-to-burst fluctuations which are very large indeed; see also \cite{chiang}. We will, in fact, conclude that from an observational point of view the relevant rates are determined by the fluctuations, and not by the average burst. Our calculations are performed by fluctuating individual bursts according to the model described above with i) the square of the distance which is assumed to follow a Euclidean or cosmological distribution, ii) with the energy assuming that the neutrino rate depends linearly on energy and follows a simple step function with ten percent of GRB producing more energy than average by a factor of ten, and one percent by a factor of 100, and, most importantly, iii) fluctuations in the $\gamma$ factor around its average value of 300. The fluctuations in $\gamma$ affect the value of $f_{\pi}$ which varies approximately as $\gamma^{-4}$ (\cite{waxman}) as well as the position of the break energy which varies as $\gamma^{2}$. For a detailed discussion see Halzen (1998). Both effects are taken into account. Clearly a factor of ten variation of $\gamma$ leads to a change in flux by roughly 4 orders of magnitude.

The origin of the large fluctuations with $\gamma$ should be a general property of boosted accelerators.  With high luminosities  emitted over short times, the large photon density renders the GRB opaque unless $\gamma$ is very large. Only transparent sources with large boost factors emit photons. They are however relatively weak neutrino sources because the actual photon target density in the fireball is diluted by the large Lorentz factor.

This raises the unanswered question whether there are bursts with lower $\gamma$-factors. Because of absorption such sources would emit less photons of lower energy and could have been missed by selection effects; they would be spectacular neutrino sources. Some have argued (\cite{stern}) for $\left<\gamma\right> \sim 30$ on the basis that the unusual fluctuations in the morphology of GRB can only be produced in a relatively dense, turbulent medium.

\section{Monte Carlo Simulation Results}

We will illustrate the effect of fluctuations by Monte Carlo simulation of GRB. For a Euclidean distribution the calculation can be performed analytically; in other cases the Monte Carlo method evaluates the integrals. The overwhelming effect of fluctuations is demonstrated in sample results shown in Figs.\,1--3 and Table\,1,a--c. Not knowing the distribution of $\gamma$-factors around its average, we have parametrized it as a Gaussian with widths $\sigma,\sigma'$ below and above the average value of 300. We chose $\sigma'$ to be either 0 or 300, to illustrate the effect of allowing GRB with Lorentz factors up to $10^3$ with a Gaussian probability. The critical, and unknown, parameter is $\sigma$. It may, in fact, be more important than any other parameter entering the calculation. As far as we know, neither theory nor experiment provide compelling information at this point. Note that we require $\sigma > 70$ in order to allow a significant part of the bursts to have $\gamma$-factors less than $10^2$, or one third the average value; see Table\,1. The value of $\sigma'$ is less critical. A value of 300 allows for Lorentz factors as large as $10^3$. The dominant effect of this is a renormalization of the neutrino rates because a fraction of the bursts now have a large $\gamma$-factor and, as a consequence, a low neutrino flux. We discuss some quantitative examples next.

Firstly, even in the absence of the dominant fluctuations in $\gamma$, the rate of 9 detected neutrinos per year over $10^3$ bursts, becomes 30--90 in the presence of fluctuations in energy and distance. The range covers a variety of assumptions for GRB distributions, which range from Euclidean to cosmological; the issue is not important here because other factors dominate the fluctuations. Every 2 years there will be an event with 7 neutrinos in a single burst. For $\sigma = 70$, the rate is $\sim600$ per year in a kilometer square detector, with 23 individual bursts yielding more than 4 neutrinos and 7 yielding more than 10 in a single year! The results for other values of $\sigma$ are tabulated in Table\,1. The number of events per year for a range of values of $\sigma$ is shown in Fig.\,1. The frequency of GRB producing seven or more neutrinos is shown in Fig.\,2. The neutrino multiplicity of bursts for $\sigma=30,\ 60$ and 75 is shown in Fig.\,3. Note that absorption in the source eventually reduces the overall rates when $\gamma$ is much smaller than average, or $\sigma$ larger than 60. This factor is included in our calculations (\cite{halzen}).

To this point we have assumed that all quantities fluctuate independently.
Interestingly, the value of $\sigma=70$ is obtained in the scenario where instead fluctuations in $\gamma$ are a consequence of fluctuations in energy. In order not to double count, fluctuations in energy itself, which only contribute a factor of 3 to the neutrino rate anyway, should be omitted.

Even though bursts with somewhat lower $\gamma$ produce neutrinos with reduced energy, compared to 700~TeV, over a longer time-scale than 1~second, the signatures of these models are spectacular. Existing detectors with effective area at the 1--10\% of a kilometer squared, should produce results relevant to the open question on the distribution of bulk Lorentz factors in the fireball model.

Independent of the numerics, the fact that a single GRB with high energy, close proximity and a relatively low Lorentz factor can reasonably produce more detectable neutrino events than all other GRB over several years time, renders the result of the straightforward diffuse flux calculation observationally misleading. Our calculations suggest that it is far more likely that neutrino detectors detect one GRB with favorable characteristics, than hundreds with average values. Clearly, our observations are relevant for other GRB models, as well as for blazars and any other boosted sources. They are also applicable to photons and may represent the underlying mechanism for the fact that the TeV extra-galactic sky consists of a few bursting sources. We expect no direct correlation between neutrino and high energy gamma sources because cosmic events with abundant neutrino production are almost certainly opaque to high energy photons.

\section{Conclusions}

Although the average event rates predicted with typical GRB parameters appear somewhat discouraging to present and future Cerenkov neutrino detector experiments, the fluctuations in these calculations are more significant and affect the prospects for detection. We speculated on the distribution of the parameters entering the fireball model calculation and used a Monte Carlo simulation to estimate the actual event rates.  The result of these simulations show that a kilometer scale detector could be expected to observe tens or hundreds of events per year. To improve on the reliability of this estimate, a well defined distribution for the Lorentz factor must be determined. Contrariwise, not observing neutrino GRB after years of observation will result in a strong limit on the number of accelerated protons in the kinetic phase of the burst, or in fine-tuned, high values of the Lorentz boost factor.

\section*{Acknowledgements}
This research was supported in part by the U.S.~Department of Energy under Grant No.~DE-FG02-95ER40896 and in part by the University of Wisconsin Research Committee with funds granted by the Wisconsin Alumni Research Foundation.
We thank G.~Barouch, T.~Piran and E.~Waxman for discussions.

\newpage

\begin{figure}
\centering\leavevmode
\epsfxsize=5in\epsffile{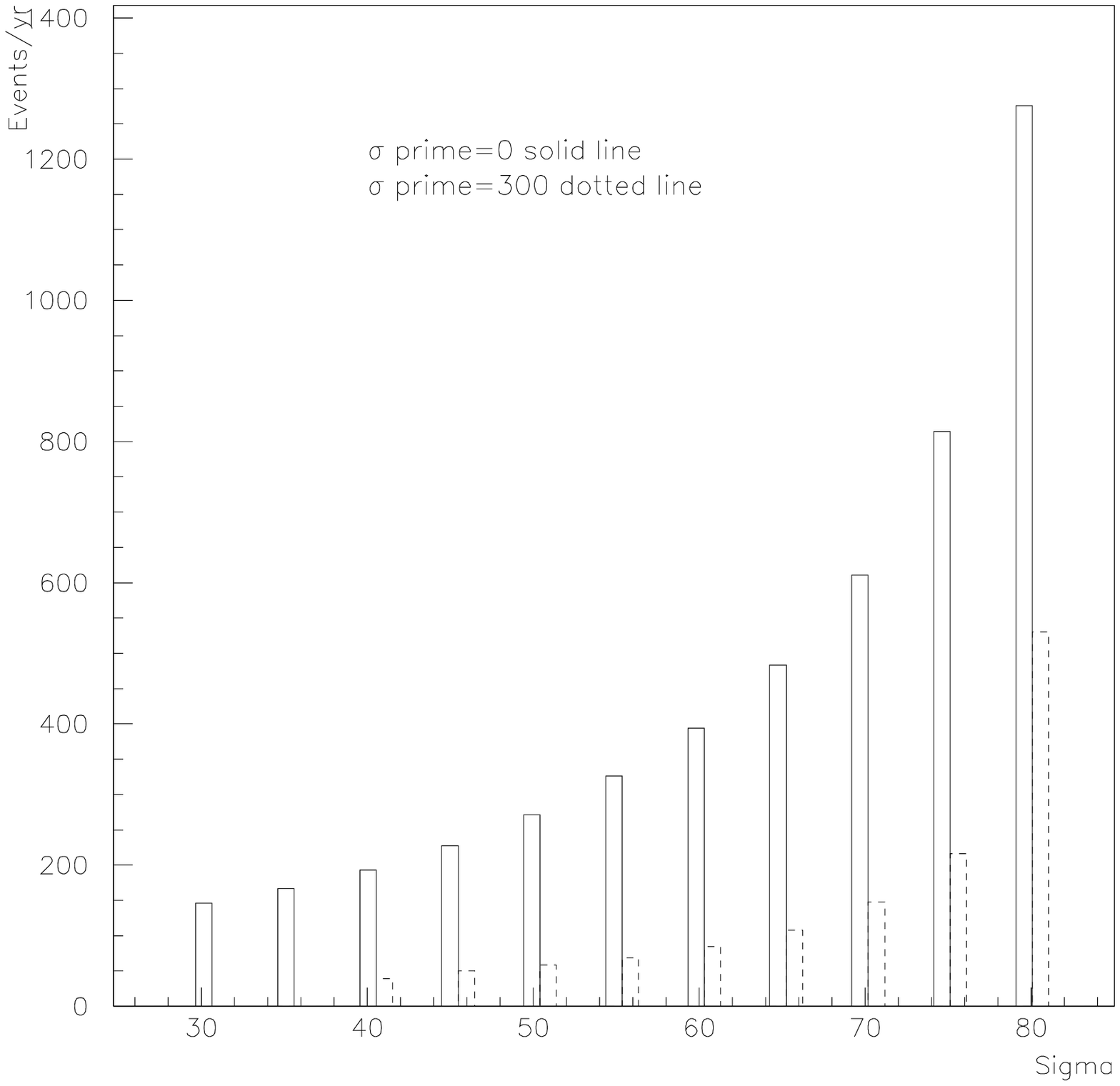}

\figcaption[fig1.eps]{Rate of high energy neutrino events in a detector with $1\rm\,km^2$ effective area as a function of the range of the bulk Lorentz factor $\gamma$. The range is assumed to be Gaussian with width $\sigma$ below the average of $300$, and $\sigma'$ above. The latter is taken to be either 0, or 300.}
\end{figure}

\begin{figure}
\centering\leavevmode
\epsfxsize=5in\epsffile{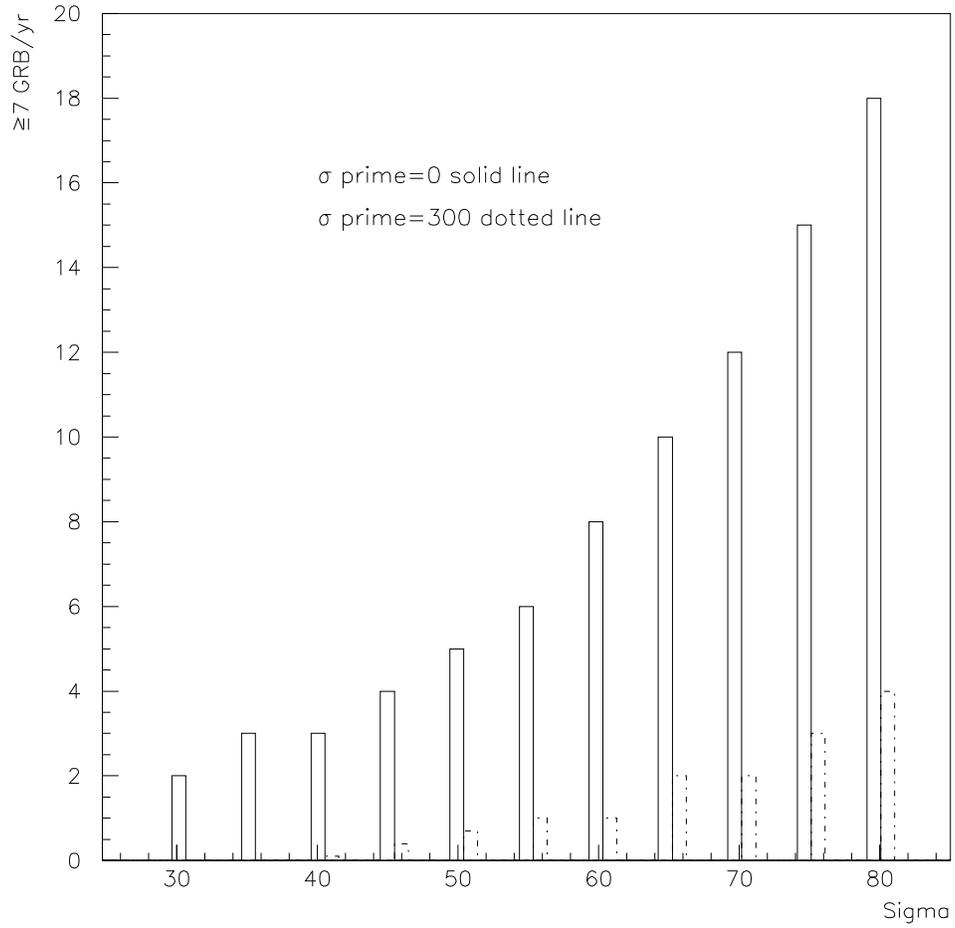}

\figcaption[fig2.eps]{Yearly rate of individual neutrino bursts with more than 7 events in a detector with $1\rm\,km^2$ effective area as a function of the range of the bulk Lorentz factor $\gamma$. The range is assumed to be Gaussian with width $\sigma$ below the average of $300$, and $\sigma'$ above. The latter is taken to be either 0, or 300.}
\end{figure}

\begin{figure}
\centering\leavevmode
\epsfxsize=5in\epsffile{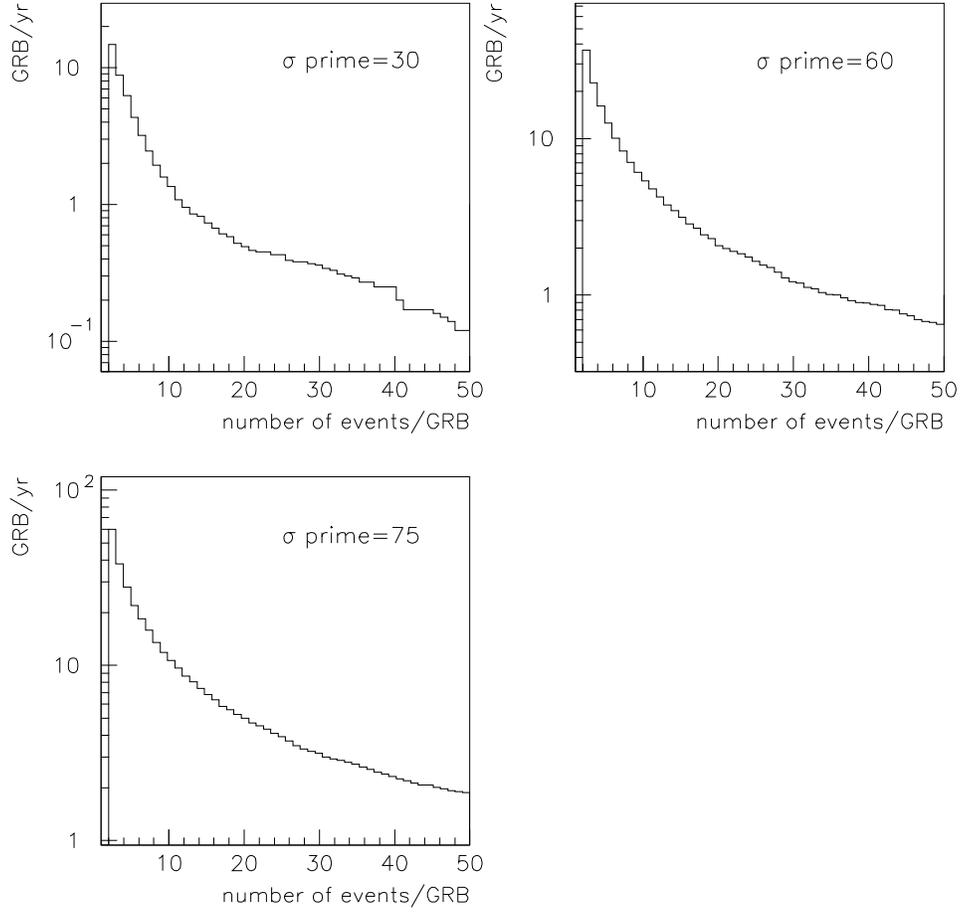}

\figcaption[fig3.eps]{Neutrino multiplicity distribution in individual GRB in a detector with $1\rm\,km^2$ effective area. The range of the bulk Lorentz factor $\gamma$ is assumed to be Gaussian with width $\sigma = 30,\ 60$ and 75 below the average of $300$, and $\sigma'$ taken to be 0.}
\end{figure}

\newpage 
\begin{table}

\caption[]{High energy neutrino events in a detector with $1\rm\,km^2$ effective area as a function of the range of the bulk Lorentz factor $\gamma$. The range is assumed to be Gaussian with width $\sigma$ below the average of $300$, and $\sigma'$ above. The latter is taken to be either 0, or 300. Shown as a function of $\sigma$ is the total number of neutrinos, as well  as the yearly number of individual bursts with more than 4, 7, and 11 high energy neutrinos. For comparison, the middle line shows the result when all fluctuations are removed.}
\bigskip
\scriptsize

\centering\leavevmode
\begin{tabular}{cccccccc@{\hspace{1.75em}}cc}
\tableline\tableline
\multicolumn{10}{c}{$\sigma'=0$}\\
\tableline
& \multicolumn{5}{c}{Percent of Lorentz Factors by Range}& &
\multicolumn{3}{l}{GRB $\geq X$ Events/yr}\\
$\sigma$& 0--10& 10--50& 50--100& 100--200& 200--300& Events/yr&
$\geq 4$& $\geq 7$& $\geq11$\\
\tableline
80& 0.04& 0.15& 1.09& 19.83& 78.9& 1276& 33& 18& 12\\
75& 0.01& 0.08& 0.68& 17.4& 81.8& 814& 27& 15& 10\\
70& 0& 0.04& 0.39& 14.8& 84.8& 611& 23& 12& 7\\
65& 0& 0.02& 0.20& 12.1& 87.7& 483& 18& 10& 6\\
60& 0& 0& 0.09& 9.46& 90.5& 394& 16& 8& 4\\
55& 0& 0& 0.04& 6.86& 93.1& 326& 13& 6& 3\\
50& 0& 0& 0.01& 4.52& 95.5& 271& 11& 5& 2\\
45& 0& 0& 0& 2.62& 97.4& 227& 9& 4& 2\\
40& 0& 0& 0& 1.28& 98.7& 193& 8& 3& 2\\
35& 0& 0& 0& 0.44& 99.6& 167& 7& 3& 1\\
30& 0& 0& 0& 0& 99.9& 146& 6& 2& 1\\
0& 0& 0& 0& 0& 100& 86& 1& 0.5& 0.2\\
\tableline\tableline
\multicolumn{10}{c}{Result if energy and distance are not allowed to vary:}\\
\tableline
0& 0& 0& 0& 0& 100& 8.6& 0& 0& 0\\
\tableline\tableline
\multicolumn{10}{c}{$\sigma'=300$}\\
\tableline
& \multicolumn{5}{c}{Percent of Lorentz Factors by Range}& &
\multicolumn{3}{l}{GRB $\geq X$ Events/yr}\\
$\sigma$& 0--10& 10--50& 50--100& 100--200& 200--300& Events/yr&
$\geq 4$& $\geq 7$& $\geq11$\\
\tableline
80& 0.01& 0.05& 0.29& 5.3& 20.8& 530& 7& 4& 3\\
75& 0& 0.03& 0.17& 4.3& 20.3& 216& 6& 3& 2\\
70& 0& 0.01& 0.10& 3.5& 19.6& 148& 5& 2& 1\\
65& 0& 0& 0.05& 2.6& 18.8& 108& 3& 2& 0.8\\
60& 0& 0& 0.02& 1.9& 17.9& 85& 3& 1& 0.7\\
55& 0& 0& 0.01& 1.2& 16.9& 69& 2& 1& 0.5\\
50& 0& 0& 0& 0.74& 15.8& 58& 2& 0.7& 0.3\\
45& 0& 0& 0& 0.39& 14.5& 50& 2& 0.4& 0.3\\
40& 0& 0& 0& 0.17& 13.1& 39& 1& 0.2& 0.1\\
\tableline\tableline
\end{tabular}
\end{table}

\end{document}